\documentclass[twocolumn,prb,aps,nobalancelastpage,amsfonts,superscriptaddress,floatfix,citeautoscript]{revtex4}

\usepackage{graphicx}
\usepackage{bm}
\begin{document}

\title{Destruction of chain-superconductivity in YBa$_2$Cu$_4$O$_8$ in a weak magnetic field}

\author{A. Serafin}
\affiliation{H.~H. Wills Physics Laboratory, University of Bristol, Tyndall Avenue, Bristol BS8 1TL, United Kingdom.}

\author{J.D. Fletcher}
\affiliation{H.~H. Wills Physics Laboratory, University of Bristol, Tyndall Avenue, Bristol BS8 1TL, United Kingdom.}

\author{S. Adachi}
\affiliation{Superconducting Research Laboratory, International Superconductivity Technology Center, Shinonome 1-10-13,
Tokyo 135, Japan.}

\author{N.E. Hussey}
\affiliation{H.~H. Wills Physics Laboratory, University of Bristol, Tyndall Avenue, Bristol BS8 1TL, United Kingdom.}

\author{A. Carrington}
\affiliation{H.~H. Wills Physics Laboratory, University of Bristol, Tyndall Avenue, Bristol BS8 1TL, United Kingdom.}

\date{\today}

\begin{abstract}
We report measurements of the temperature dependent components of the magnetic penetration depth $\lambda(T)$ in single
crystal samples of YBa$_2$Cu$_4$O$_8$ using a radio frequency tunnel diode oscillator technique. We observe a downturn
in $\lambda(T)$ at low temperatures for currents flowing along the $b$ and $c$ axes but not along the $a$ axis.  The
downturn in $\lambda_b$ is suppressed by a small dc field of $\sim 0.25$\,T.  This and the zero field anisotropy of
$\lambda(T)$ likely result from proximity induced superconducting on the CuO chains, however we also discuss the
possibility that a significant part of the anisotropy might originate from the CuO$_2$ planes.
\end{abstract}
\maketitle

\begin{figure*}
\center
\includegraphics*[width=16cm]{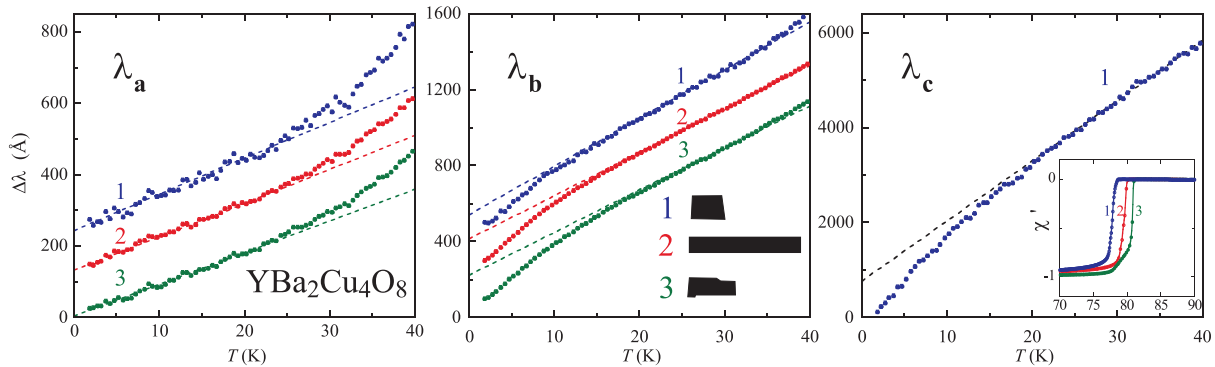}
\caption{(color online) Temperature dependence of the penetration depth for the three principle directions for
YBa$_2$Cu$_4$O$_8$. The shapes of the samples are shown in the $\lambda_b$ panel. Their dimensions ($\ell_a \times
\ell_b \times \ell_c$) are: sample 1: $115\times 80 \times 15 \mu$m$^3$, sample 2: $60 \times 420 \times 8 \mu$m$^3$,
sample 3: $170\times60\times6 \mu$m$^3$. The RF susceptibility ($\chi^\prime$) close to $T_c$ (normalized to $-1$ at
low temperature) is shown in the $\lambda_c$ panel.}\label{Fig:lambda}
\end{figure*}

The Y-based high-$T_c$ cuprate superconductors are unique in that they have quasi-one-dimensional CuO chain structures
in addition to the CuO$_2$ planes in which the interactions responsible for superconductivity are thought to originate.
The effect of these chain layers on the normal and superconducting state properties has long been debated.  Indeed,
since the Y-based cuprates are so far the only hole-doped cuprates which show quantum oscillations in their underdoped
state \cite{Doiron-leyraudPLLBLBHT07,BanguraFCLNVHDLTAPH08,YellandSMHBDC08}, it is natural to question if the coherent
electron orbits could in fact originate in the chain parts of the Fermi surface (FS)
\cite{CarringtonY07,Riggs1008.1568}. A important question is whether the chain FS is superconducting and if so is this
superconductivity quenched by a much smaller magnetic field than the plane.

YBa$_2$Cu$_3$O$_{7-\delta}$ (Y-123) contains a single CuO chain per unit cell whose oxygen content can be varied to
tune the doping level on the CuO$_2$ plane. Its close relative YBa$_2$Cu$_4$O$_{8}$ (Y-124) has a double chain layer
that is stoichiometric and a planar state that is slightly underdoped \cite{BanguraFCLNVHDLTAPH08}. According to
band-structure calculations \cite{CarringtonY07,AmbroschdraxlBS91,YuPF91} the chains form quasi-1D electronic bands
which cut across the CuO$_2$ plane FS sheets, with considerable hybridization between the states close to the crossing
points. As the electron-pairing interaction which gives rise to superconductivity in the cuprates is likely to rely
strongly on the electronic structure of the quasi-two-dimensional plane, it might seem unlikely that there any
intrinsic pairing of electrons on the chain. Indeed, in isostructural Pr-124, the double chain network is metallic yet
exhibits no superconductivity down to 0.5\,K \cite{HusseyMBBHI02}.

In both Y-123 and Y-124, the CuO chains have been modelled as an essential normal layer which is coupled to the planes
via single electron tunnelling, similar to the classical proximity effect between normal metals and superconductors
\cite{AtkinsonC95,XiangW96}. This is augmented by the strong hybridization between plane and chain states which occurs
at certain momentum values.  This model predicts that the chain states will have low energy gap structures arising from
variations of the gap within the chain FS.  This implies that the superfluid density should have quite different
temperature dependencies for screening currents flowing in the $a$ or $b$ (chain) directions. Experimentally however,
it is found that for Y-123 the temperature dependence of $\lambda_a$ and $\lambda_b$ are very similar
\cite{ZhangBKLBHBT94,CarringtonGKG99} although their zero temperature values differ \cite{BasovLBHDQTRGT95,Kiefl10}.
This has led to the proposal that there are intrinsic pairing interactions on the chains and that the planes and chains
are predominately coupled by Josephson-like pair tunnelling \cite{XiangW96}.  Alternatively, this has been explained by
chain disorder \cite{Atkinson99}.

In this paper, we report measurements of the anisotropic components of the magnetic penetration depth in Y-124 single
crystals as a function of temperature and dc field. A rapid suppression of the superconducting component on the CuO
double chain is observed in a small applied field. These results are significantly different to those obtained on
optimally doped Y-123 and appear to support the proximity effect model for Y-124. However, some questions remain.

Single crystals of YBa$_2$Cu$_4$O$_{8}$ were grown using a high oxygen pressure flux based method
\cite{AdachiNTNTHI98}. Penetration depth was measured using a radio frequency (RF) tunnel diode oscillator method
\cite{CarringtonGKG99} operating at $\sim$ 12\,MHz. The sample is attached with vacuum grease to a sapphire rod and is
placed in a copper coil which forms part of the oscillator's tank circuit.  A small superconducting solenoid allows us
to apply a dc field co-linear with the weak RF probe field ($H_{\rm RF}\simeq 10^{-6}$\,T).

Changes in the resonant frequency of our RF oscillator $\Delta F$ are directly proportional to changes in sample's
superconducting volume as a function of temperature or field.  The samples are thin platelets with dimensions
$\ell_{a,b,c}$ along the respective crystallographic directions (Fig.\ \ref{Fig:lambda}).  With the RF field applied
along the $b$ direction the change in frequency $\Delta F$ due to changes in the $\lambda_a$ and $\lambda_c$ is given
by $\Delta F=2\alpha\beta (\ell_a \ell_b \Delta\lambda_a + \ell_b \ell_c \Delta\lambda_c)$ where $\alpha$ is a constant
set by the coil geometry and $\beta$ is the effective sample demagnetizing factor (here we neglect the small
contribution from the currents on small sample surfaces perpendicular to the field).  Hence, contributions from two
components of $\lambda$ are always mixed.   In principle, it is possible to separate the components by making
measurements of $\Delta\lambda$ with $H_{\rm RF}$ in all three directions, however in practice this approach is
inaccurate because of uncertainties in the demagnetizing factor particularly in the $H\|c$ configuration.  Instead, we
cleaved the sample (using a razor blade) halving $\ell_a$ and thus doubling the contribution of $\Delta\lambda_c$
\cite{BonnKBLHHBT96}. Measurements of sample 1 with $H\|b$ (where the demagnetising effect is small) before and after
cleaving thus allow us to isolate the $\lambda_c$ contribution which then can be subtracted from the measurements with
$H\|a$ and $H\|b$ to yield the three components of $\lambda$. Consistency was checked by cleaving the sample a second
time.  The extracted $\Delta \lambda_c(T)$ was used to extract $\Delta \lambda_a$ and $\Delta \lambda_b$ from two
further samples.

The temperature dependence of the three extracted components of $\lambda$ below $T$ = 40 K are shown in Fig.\
\ref{Fig:lambda}. The data for $\Delta \lambda_a(T)$ and $\Delta \lambda_b(T)$ for all three samples is very
consistent. $\lambda_a$ has a linear $T$ dependence from base temperature up to $\sim 15$\,K, with slope
$d\lambda_a/dT=9.6(6)$\AA/K which is consistent with a simple $d$-wave model.  The behavior of $\Delta \lambda_b$ and
$\Delta \lambda_c$ are quite different to $\Delta \lambda_a$. Between $\sim 15$ and 30\,K $\Delta \lambda_b(T)$ is
linear with a slope $d\lambda_b/dT=24(2)$\AA/K which is $\sim$2.5 times larger than for $\lambda_a$. Below $\sim$15\,K
however, there is a downturn in $\Delta\lambda_b(T)$ which indicates the onset of  additional screening which reduces
$\lambda_b$ by $\sim$150\AA~ at $T=2$\,K compared to the extrapolated linear behavior.  Similar behavior is found for
$\Delta\lambda_c$ where the downturn sets in at approximately the same temperature and $d\lambda_c/dT=130$\AA/K.   A
downturn in $\lambda(T)$ had been seen in previous measurements of Y-124 \cite{Manzano02,LamuraGKKA06} however the
separate contributions of $\lambda_a$ and $\lambda_b$ were not determined.

In the conventional model of $d$-wave superconductivity \cite{XuYS95} with a circular 2D FS we expect
$\frac{d\lambda}{dT}=\frac{2\ln 2 \lambda(0)}{d\Delta/d\phi |_{\rm node}}$.  We cannot measure $\lambda(0)$ directly in
our experiments. However, infra-red reflectivity (IRR) experiments \cite{BasovLBHDQTRGT95} on Y-124 give
$\lambda_a(0)=2000$\,\AA~ and $\lambda_b(0)=800$\,\AA,  (the anisotropy $\lambda_a/\lambda_b = 2.5$).  Although the
$\lambda$ anisotropy measured by IRR for Y-123 ($\lambda_a/\lambda_b \simeq 1.6$) is somewhat higher than that measured
by other other techniques ($\lambda_a/\lambda_b \simeq 1.2$) \cite{Kiefl10} it does seem likely that
$\lambda_a/\lambda_b$ is significantly greater than unity for Y-124 which is in the opposite direction to our measured
anisotropy in $d\lambda/dT$.  Hence, Fermi velocity anisotropy alone cannot explain the data within this simple model.
If the order parameter had a significant $s$ component (i.e., $d + \eta s$) so that the nodes move away from
$\phi=45^\circ$ towards the $b$-direction, then the paramagnetic current produced by the thermally excited nodal
quasiparticles could be larger in the $b$-direction and this, in principle, could overcome the opposite anisotropy of
$\lambda(0)$.  Within a single band model a value of $\lambda_a/\lambda_b = 2.5$ implies a substantial anisotropy of
the in-plane Fermi velocity. Calculations based on fits to photoemission results \cite{Kondo09} however, suggest that
at a doping level of $p=0.14$ the anisotropy of the in-plane superfluid density is only $\sim 2$\%.

\begin{figure} \center
\includegraphics*[width=7cm]{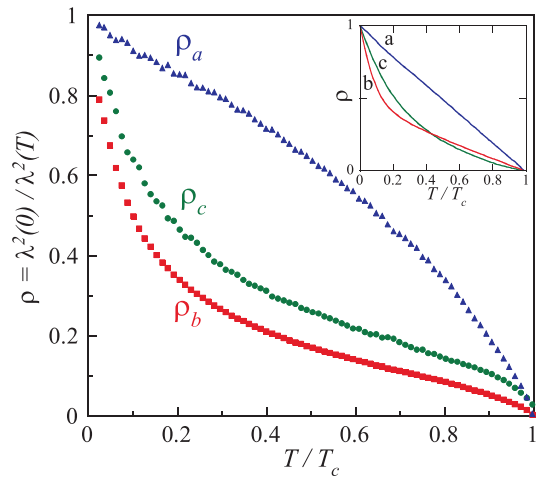}
\caption{(color online) Normalized superfluid density [$\rho=(\lambda(0)^2/(\lambda(0)+\Delta\lambda(T))^2$] for sample
2 in all three directions using values of $\lambda(0)$ taken from infra-red measurements \cite{BasovLBHDQTRGT95}
($\lambda_a(0)=2000$\,\AA~and $\lambda_b(0)=$\,800\AA) and aligned polycrystalline measurements
($\lambda_c(0)=6150$\,\AA) \cite{PanagopoulosTX99}. The inset shows theoretical predictions for $\rho$ in the proximity
coupling (single electron tunneling) model taken from Ref.\ \onlinecite{XiangW96}.}\label{Fig:superfluid}
\end{figure}

An explanation for both the larger value of $\lambda(0)$ and $\Delta\lambda(T)$ in the $b$-direction can be found in
the plane-chain proximity models mentioned above.  In Fig.\ \ref{Fig:superfluid} we show the normalized superfluid
density $\rho=\lambda(0)^2/(\lambda(0)+\Delta\lambda(T))^2$ for sample 2, calculated using the values of $\lambda(0)$
taken from infra-red measurements \cite{BasovLBHDQTRGT95} ($\lambda_a$ and $\lambda_b$) and aligned polycrystalline
measurements ($\lambda_c$) \cite{PanagopoulosTX99}.  The behavior in the $a$-direction is similar to that found for
$\rho_a$ and $\rho_b$ of Y-123, and varies linearly with $T$ up to $\sim T_c/2$. However, both $\rho_b(T)$ and
$\rho_c(T)$ are quite different to $\rho_a(T)$ with both showing strong upward curvature for temperature below $\sim
T_c/3$.  Note that the small downturn in $\Delta\lambda_b(T)$ below 15\,K noted above (Fig.\ \ref{Fig:lambda}) is not
the primary reason for the disparity between $\rho_a$ and $\rho_b$ in Y-124; rather it is caused by the substantial
difference in the ratio $(d\lambda/dT)/\lambda(0)$ which is $4.8\times 10^{-3}$\,K$^{-1}$ for the $a$ axis but $\sim 6$
times larger ($30\times 10^{-3}$\,$K^{-1}$) for the $b$ axis.  Clearly small uncertainties in the values of
$\lambda(0)$ will not have a large effect on this.  These upturns are similar to those reported for the average
in-plane and $c$ axis superfluid density measured on polycrystalline samples \cite{PanagopoulosTX99}.

In the inset to Fig.\ \ref{Fig:superfluid} we show the predictions of the single electron tunnelling plane-chain
proximity coupling model of Ref.\ \onlinecite{XiangW96} (similar results using a similar model are also given in Ref.\
\onlinecite{AtkinsonC95}).  These predictions are very similar to our experimental findings, with both $\rho_b(T)$ and
$\rho_c(T)$ having strong upward curvature below $\sim T_c/3$.  In this particular simulation both plane and chain FS
were assumed to be superconducting, but similar results are found when there is no intrinsic gap on the chains
\cite{AtkinsonC95}. A feature of this proximity model is that the upward curvature of $\rho_b$ is strongly suppressed
by impurity scattering (e.g. broken chain segments) \cite{Atkinson99}.  The difference between Y-123 and Y-124 may then
be due to fact that Y-124 has completely full chains whereas in Y-123 the chain segments are much shorter (more
disorder). Inter-chain coupling in Y-124 will reduce localization effects which are likely to be present in the more 1D
Y-123 chains.

\begin{figure}[b]
\center
\includegraphics*[width=7cm]{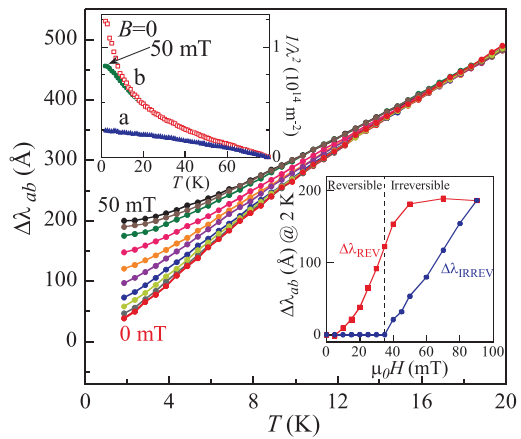}
\caption{(color online) Field dependence of $\lambda_{ab}$ for sample 2 (dimensions $60 \times 300 \times 8 \mu$m$^3$)
from 0 to 50\,mT in 5\,mT increments. The $b$-axis dimension is less than that stated in Fig.\ \ref{Fig:lambda} because
it was broken between runs. The aspect ratio of this sample means that $\sim$80\% of $\lambda_{ab}$ comes from
$\lambda_b$. To remove artifacts arising from the field dependent (but temperature independent) background from the
measurement coil the data have been shifted in frequency so that they coincide at $T=20$\,K for all fields.  The lower
inset shows the change in $\lambda$ at the fixed temperature of 2\,K along with the irreversible changes to $\lambda$,
showing the onset of flux entry at $\sim 35$\,mT. The upper inset shows the superfluid density along $a$ and $b$ axes
in zero field together with the $b$ axis data in 50\,mT. Here $\Delta \lambda_b(T)$ at 50\,mT has been approximated by
$1.2\Delta\lambda_{ab}$ . The factor 1.2 accounts for the small contribution of $\lambda_a$ to $\lambda_{ab}$.}
\label{Fig:fielddependence}
\end{figure}

Next we discuss the field dependence of $\lambda$.  Here we apply a dc field co-linear with the RF probe field along
the $c$ direction where $H_{c1}$ is maximal. Data for sample 2 are shown in Fig.\ \ref{Fig:fielddependence} for dc
fields up to 50\,mT. In this field configuration the measured $\Delta \lambda_{ab}$ is a mixture of $\Delta \lambda_a$
and $\Delta\lambda_b$ but because of the aspect ratio of this particular sample ($\ell_b/\ell_a\simeq 5$) around 80\%
of $\Delta \lambda_{ab}$ comes from $\Delta\lambda_b$.  In this configuration the calibration factor is difficult to
calculate (especially for a sample with such an elongated shape), so we determined it experimentally by measuring, in
both the $H\|ab$ and $H\|c$ configurations, a twinned sample of Y-123 with almost identical shape.  From the total
frequency shifts measured when the sample was withdrawn from the coil and from the sample dimensions we estimate that
the demagnetizing field enhancement factor $\beta\simeq 5$, hence the effective surface fields are 5 times larger than
the applied field.

For $H_{\rm dc}=0$  (Fig.\ \ref{Fig:lambda}) a strong downturn in $\lambda_{ab}(T)$ is seen below $T\simeq$\,15K as in
the direct measurements of $\Delta\lambda_b(T)$ (Fig.\ \ref{Fig:lambda}).  As $H_{\rm dc}$ is increased, initially
there is no change and then at $\mu_0H_{\rm dc}\simeq 5$\,mT, $\lambda_{b}(T=2$\,K$)$ starts to increase  monotonically
and eventually saturates for $\mu_0H_{\rm dc}\gtrsim 50$\,mT (see inset Fig.\ \ref{Fig:fielddependence}).  The
temperature dependence of $\lambda_{ab}$ also changes, from $\sim T^{0.7}$ for $H_{\rm dc}=0$ to $\sim T^{1.2}$ for
$\mu_0H_{\rm dc}=50$\,mT.  The applied dc field wipes out the downturn in $\lambda_{ab}$ and returns the temperature
dependence to a quasi-linear variation expected from a conventional $d$-wave superconductor with small amounts of
impurities \cite{Hirschfeld93}. Similar behavior with comparable field scales was observed in two other samples. The
high field behavior of $\lambda_{ab}(T)$ is less linear than $\lambda_a$ at the lowest temperatures perhaps because of
anisotropic impurity scattering.

For  $\mu_0H_{\rm dc}\lesssim 35$\,mT $\Delta \lambda_{ab}$ is perfectly reversible on cycling from base temperature
(where the field was increased) up to 20\,K and back again.  For fields above this $\Delta\lambda_{ab}(T)$ was
hysteretic on the first temperature cycle and then reversible thereafter. This irreversibility increased linearly with
field (see inset Fig.\ \ref{Fig:fielddependence}).  We attribute this hysteresis to vortex motion.  Initially, as the
vortices enter the sample they are not in deep pinning wells and so give a strong contribution to the measured
effective $\lambda$. Upon temperature cycling they settle on strong pinning sites (or move to the center of the sample)
and therefore no longer give a strong contribution to $\lambda$.  As the field at which $\lambda_{ab}(T)$ begins to
change is much below the field at which the hysteretic behavior set in, and since for fields greater than this the
reversible part of $\Delta\lambda(H)$ begins to saturate whereas the irreversible part increases linearly with $H$, it
is clear that vortex motion is not responsible for the changes in the reversible part of $\Delta\lambda_{ab}(H)$ which
are shown in the main part of Fig.\ \ref{Fig:fielddependence}.  Rather reversible changes reflect a decrease in the
Meissner screening current. Direct measurements of the dc magnetization using a SQUID magnetometer showed that the
field of first flux penetration was $\sim$35\,mT, for this sample and field configuration at $T=10$\,K, which coincides
with the onset of irreversibility.

Field dependence of $\lambda$ in the Meissner state can result from several effects. A linear increase of $\lambda$
with $H$ is expected from Doppler shifted quasiparticles close to the nodal points \cite{XuYS95}. This effect is
however, very small ($\sim$2\,\AA~for $\mu_0H=10$\,mT for Y-123) and has never been conclusively observed (the measured
$\lambda(H)$ in Y-123 does not have the temperature or field orientation dependence expected from theory and may
therefore have another origin) \cite{CarringtonGKG99,BidinostiHBL99}.  A decrease in  $\lambda$  with increasing field
can also result from the presence of Andreev bound states,\cite{CarringtonMPGKT01} but this is in the opposite
direction to what is observed here.

Instead, the most likely explanation is that in Y-124 the small dc field quenches a least part of the superconducting
current in the chains. We might expect this to occur when the Doppler shift of the chain quasiparticle energies are of
order the chain gap.  It is often observed that proximity-driven superconductivity is quenched in small fields
\cite{ProzorovGBC01}, for example the $\pi$ band superconductivity in MgB$_2$ \cite{Bouquet02}. In Y-124 the magnetic
field does not affect $\lambda(T)$ above $\sim 15$\,K and the strong anisotropy in $d\lambda/dT$ between the $a$ and
$b$ direction as discussed above remains.  Also the field induced change in $\lambda_{ab}\simeq 190$\,\AA~ at the
lowest temperature which does not significantly change the anisotropy of $\lambda$. This is illustrated in the inset to
Fig.\ \ref{Fig:fielddependence} where we show the field suppressed superfluid density along the $b$ axis along with the
zero field results.

The fact that only part of the excess $b$ axis superfluid (relative to the $a$ axis) is suppressed in weak field
($\sim$0.25\,T including demagnetizing effects) might suggest that there is significant variation of the gap along the
chain FS \cite{Atkinson09}. If the system is sufficiently clean, so that scattering between regions of the chain FS
with different gaps (resulting from a variation of the plane-chain hybridization) is small, then the downturn in
$\lambda_b(T)$ and $\lambda_c(T)$ for $T\lesssim$15\,K could result from a region of chain FS with weak pairing
becoming superconducting. This region then becomes normal when a weak field is applied. Alternatively, the downturn and
field dependence could reflect the superconductivity in the chain as whole, and then the remaining anisotropy in high
field would reflect the anisotropy of the CuO$_2$ planes. Although, as mentioned above, there is no indication of such
anisotropy from photoemission experiments we note that quantum oscillation \cite{BanguraFCLNVHDLTAPH08,YellandSMHBDC08}
and magnetotransport \cite{RourkeBPLDLTASH10} measurements
 show that there is a significant reconstruction of the FS in this material at low temperature.
 Recently, evidence of significant electronic anisotropy in the CuO$_2$ planes of underdoped Y-123 has been found in Nernst effect
 measurements \cite{DaouCLCLDRLBHT10} and it is likely that properties of Y-124 are similar.  Further theoretical work calculating the effect of magnetic field on the superfluid
density in the proximity models \cite{AtkinsonC95,XiangW96} would help decide between these two competing
interpretations. Completing the picture of plane-chain superconductivity in Y-124 and Y-123 should be help our
understanding of quantum oscillations and the underlying electronic structure of these materials.

We thank Francisco Manzano and Ruslan Prozorov for their contributions to this project and David Broun and Bill
Atkinson for useful discussions. This work was funded by the UK EPSRC.
\bibliography{Y124_lambdaT}

\bibliographystyle{aps5etal}

\end{document}